\documentclass{aa}
\usepackage{graphicx}
\usepackage{txfonts}
\usepackage{lipsum}
\usepackage{subcaption}
\usepackage{lscape}
\usepackage{placeins}
\usepackage[hidelinks]{hyperref}
\begin{document}

   \title{Search for L4 Earth Trojan asteroids with the 2.5-meter Wide Field Survey Telescope}

   \author{Junqiang Lu \inst{1,2} \corrauth{ljq1999@mail.ustc.edu.cn}
        \and Lulu Fan \inst{1,2,3,5} \corrauth{llfan@ustc.edu.cn}
        \and Shaohan Wang \inst{1,2} \email{wangshaohan@mail.ustc.edu.cn}
        \and Minxuan Cai \inst{1,2} \email{caimx@mail.ustc.edu.cn}
        \and Bingxue Fu \inst{1,2} \email{fbx001128@mail.ustc.edu.cn}
        \and Xu Kong \inst{1,2,3} \email{xkong@ustc.edu.cn}
        \and Haibin Zhao \inst{4} \email{meteorzh@pmo.ac.cn}
        \and Bin Li \inst{4} \email{binli@pmo.ac.cn}
        \and Qingfeng Zhu \inst{1,2} \email{zhuqf@ustc.edu.cn}
        \and Zhen Wan \inst{1,2} \email{zhen_wan@ustc.edu.cn}
        \and Feng Li \inst{6} \email{phonelee@ustc.edu.cn}
        \and Ming Liang \inst{7} \email{liangming@gmail.com}
        \and Binyang Liu \inst{4,2} \email{byliu@pmo.ac.cn}
        \and Zheng Lou \inst{4,2} \email{zhenglou@pmo.ac.cn}
        \and Jinlong Tang \inst{8} \email{ioetang@163.com}
        \and Hairen Wang \inst{4,2} \email{hairenwang@pmo.ac.cn}
        \and Jian Wang \inst{6,3} \email{wangjian@ustc.edu.cn}
        \and Yongquan Xue \inst{1,2} \email{xuey@ustc.edu.cn}
        \and Hongfei Zhang \inst{6} \email{nghong@ustc.edu.cn}
    }

   \institute{Department of Astronomy, University of Science and Technology of China, Hefei 230026, China
   \and School of Astronomy and Space Science, University of Science and Technology of China, Hefei 230026, China
   \and Institute of Deep Space Sciences, Deep Space Exploration Laboratory, Hefei 230026, China
   \and Purple Mountain Observatory, Chinese Academy of Sciences, Nanjing 210023, China
   \and College of Physics, Guizhou University, Guiyang, Guizhou 550025, China
   \and State Key Laboratory of Particle Detection and Electronics, University of Science and Technology of China, Hefei 230026, China
   \and National Optical Astronomy Observatory (NSF’s National Optical-Infrared Astronomy Research Laboratory) 950 N Cherry Ave. Tucson Arizona 85726, USA
   \and Institute of Optics and Electronics, Chinese Academy of Sciences, Chengdu 610209, China
   }

\abstract
   {Earth Trojan asteroids (ETAs) are a mysterious population, and dynamically stable ETAs—if primordial—could be "living fossils" of the early solar system. To date, there are only two known ETAs, but both are temporary ETAs.}
   {The aim of our survey is to discover new temporary or stable ETAs; in the absence of detections, we derive upper limits on the population of stable ETAs.} 
   {We conducted the largest wide-area survey of the Earth's L4 Lagrange point region so far using the Wide Field Survey Telescope, covering about 236.74~deg$^2$, corresponding to 33.24\% of the probability coverage for sky regions where dynamically stable L4 ETAs are likely to reside.}
   {No new ETAs were detected in our survey. We place a cumulative upper limit of $N(H<19.1) \lesssim 19$ on the stable population of objects larger than $\sim$520 m (for an assumed albedo of 0.15). This represents the most stringent constraint on the ETA population to date.}
   {}


\maketitle
\nolinenumbers

\section{Introduction}\label{Introduction}

Earth Trojan asteroids (ETAs) are a poorly understood population of small bodies trapped in the 1:1 mean-motion resonance with Earth, residing near the L4 or L5 Lagrange points. 
Some ETAs may be primordial remnants from the protoplanetary disk, potentially as old as the Solar System itself \citep{cukLongtermStabilityHorseshoe2012}. As such, they could serve as ``living fossils,'' offering valuable insights into the formation and early evolution of our planetary system. 
Notably, the leading hypothesis for the origin of the Moon invokes a giant impact between the proto-Earth and a Mars-sized body, Theia, which may have originated as an ETA \citep{belbrunoWhereDidMoon2005}. 
Furthermore, the ETA population has been proposed to explain the observed asymmetry in crater distribution between the Moon’s leading and trailing hemispheres \citep{itoAsymmetricImpactingMoon2010}. 
Due to their low relative velocities with respect to Earth \citep{malhotraCaseDeepSearch2019a}, ETAs represent favorable targets for future space missions. 
Additionally, as members of the near-Earth object (NEO) population, some ETAs may pose long-term impact hazards to terrestrial life and infrastructure. 
Consequently, systematic discovery and monitoring efforts are both scientifically valuable and relevant to planetary defense.

Numerical simulations and analytical studies have been used to explore the dynamical properties of ETAs. 
\citet{tabachnikAsteroidsInnerSolar2000} identified two stable regions for ETAs: one at low inclinations and another at moderate inclinations.
\citet{dvorakOrbit2010TK72012a} obtained similar results with slightly different inclination ranges and discovered an additional small, U-shaped stable window near $i \approx 50^\circ$.
\citet{marzariLongTermStability2013a} investigated the effects of the Earth–Moon tidal forces and the Yarkovsky effect on ETAs, finding that they can survive for timescales of $\sim 10^9$~yr, although their orbital evolution is chaotic.
\citet{zhouOrbitalStabilityEarth2019a} presented detailed dynamical maps in the $(a_0, i_0)$ plane and ruled out the existence of small primordial ETAs.
Using large-scale numerical integrations, \citet{yeagerMEGASIMLifetimesResonances2022} constructed synthetic ETA populations and predicted complete depletion by 2.33~Gyr.
In a follow-up study, \citet{yeagerMEGASIMDistributionDetection2024} used the same datasets to assess the detectability of ETAs under the observational strategies of the Zwicky Transient Facility and the Vera C.\ Rubin Observatory’s Legacy Survey of Space and Time.

Observational searches for ETAs have so far yielded stringent upper limits but few detections. 
On 1994 May 5–7 and July 6–8 UT, \citet{whiteleyCCDSearchLagrangian1998} used the University of Hawaii 2.24-m telescope to cover 0.35~deg$^2$ around the L5 point and found no ETA candidates, deriving the first crude upper limit of $\sim$3 objects per square degree.
\citet{cambioniUpperLimitEarths2018} conducted an L4 survey using the OSIRIS-REx spacecraft during its cruise to asteroid Bennu, covering nine $4^\circ \times 4^\circ$ fields and establishing an upper limit of $73 \pm 23$ ETAs at absolute magnitude $H = 20.5$.
\citet{yoshikawaMissionStatusHayabusa22018a} searched for ETAs near L5 with the Hayabusa2 spacecraft en route to asteroid Ryugu but reported no candidates and provided no quantitative upper limit.
\citet{markwardtSearchL5Earth2020} used the Dark Energy Camera (DECam) to cover 24~deg$^2$ near L5, setting a more stringent upper limit.
\citet{lifsetSearchL4Earth2021} employed DECam to survey 5.72~deg$^2$ near L4—covering a smaller area than \citet{markwardtSearchL5Earth2020} but reaching deeper limiting magnitudes—and derived updated upper limits on the ETA population by incorporating modern asteroid population models and extensive numerical simulations.
Despite these dedicated searches yielding null results, two ETAs have been serendipitously discovered:  
(706765)~2010~TK$_7$ \citep{connorsEarthsTrojanAsteroid2011} and (614689)~2020~XL$_5$ \citep{huiSecondEarthTrojan2021}.  
Both objects librate around the L4 Lagrange point; however, dynamical analyses indicate that they are transient co-orbitals, with estimated residence times in the Trojan configuration of only a few thousand to tens of thousands of years—far shorter than the age of the Solar System.

\section{Data}
\subsection{WFST Observations}

The Wide Field Survey Telescope \citep[WFST]{wangSciences25meterWide2023} is a 2.5-m telescope located in Lenghu, Qinghai Province, northwestern China, jointly developed by the University of Science and Technology of China and the Purple Mountain Observatory.  
It features a circular field of view with a diameter of 3$^\circ$, a camera composed of nine $9.2\,\mathrm{K} \times 9.2\,\mathrm{K}$ CCD detectors, and six broadband filters ($u$, $g$, $r$, $i$, $z$, $w$).\footnote{\url{wfst.ustc.edu.cn/}} 
WFST is expected to be a powerful facility for time-domain astronomy, with discovery potential for a wide range of transients and moving objects. Key science targets include tidal disruption events \citep{linProspectsFindingTidal2022}, active galactic nuclei \citep{suCouldInterbandLag2024}, kilonovae \citep{liuTargetofOpportunityObservationDetectability2023}, Type Ia supernovae \citep{huProspectsSearchingType2022}, globular clusters \citep{wanTidalStructuresSix2025}, variable stars \citep{linMinutecadenceObservationsGalactic2025a}, and asteroids \citep{luSearchCapabilityNearEarth2025,wangHeliocentricorbitingObjectsProcessing2025}.

\begin{figure*}[ht!]
   \centering
   \includegraphics[width=\textwidth]{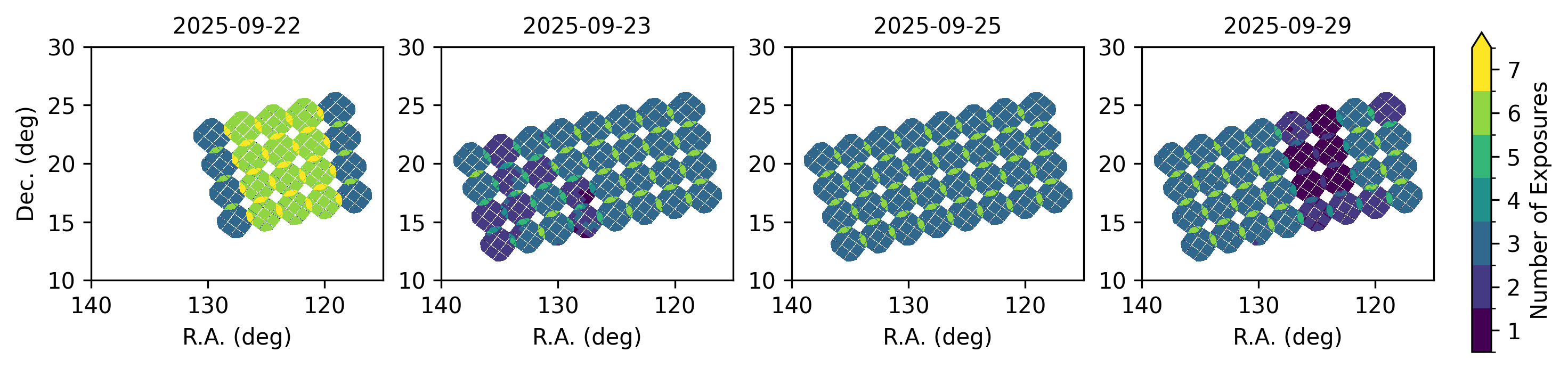}
   \caption{Sky coverage of the WFST L4 ETAs survey. The color scale indicates the number of usable exposures per sky position (in equatorial coordinates). 
   Each pointing exhibits small gaps between the nine CCDs, and the four corners of the CCD array are truncated due to being outside the effective imaging area.
   Due to an error in the observing plan provided for the night of 2025 September 22, only 20 of the 32 intended pointings were executed; among these, 12 were repeated six times instead of the nominal three.
   Observations on 2025 September 22 and 25 were conducted under good weather conditions, yielding fully usable data. In contrast, poor weather on 2025 September 23 and 29 resulted in partial data loss for those nights.}
   \label{fig:footprints_radec}
\end{figure*}

\begin{table}[htb!]
\centering
\caption{WFST pointings in the L4 ETAs survey.}
\label{tab:Targets_radec}
\begin{tabular}{clcc}
\hline
\hline
Group & Target & R.A. & Dec.\\
\hline
1 & p1 & 119.03404° & 24.58216° \\
1 & p9 & 118.51207° & 22.12851° \\
1 & p17 & 118.00799° & 19.67325° \\
1 & p25 & 117.51913° & 17.21663° \\
1 & p2 & 121.71572° & 24.07934° \\
1 & p10 & 121.15228° & 21.63383° \\
1 & p18 & 120.60762° & 19.18649° \\
1 & p26 & 120.07893° & 16.73759° \\
2 & p3 & 124.37726° & 23.53707° \\
2 & p11 & 123.77429° & 21.10008° \\
2 & p19 & 123.19082° & 18.66102° \\
2 & p27 & 122.62390° & 16.22022° \\
2 & p4 & 127.01793° & 22.95683° \\
2 & p12 & 126.37743° & 20.52864° \\
2 & p20 & 125.75696° & 18.09816° \\
2 & p28 & 125.15348° & 15.66576° \\
3 & p5 & 129.63718° & 22.34015° \\
3 & p13 & 128.96118° & 19.92096° \\
3 & p21 & 128.30557° & 17.49929° \\
3 & p29 & 127.66723° & 15.07552° \\
3 & p6 & 132.23467° & 21.68861° \\
3 & p14 & 131.52522° & 19.27855° \\
3 & p22 & 130.83636° & 16.86583° \\
3 & p30 & 130.16489° & 14.45086° \\
4 & p7 & 134.81023° & 21.00383° \\
4 & p15 & 134.06938° & 18.60295° \\
4 & p23 & 133.34916° & 16.19925° \\
4 & p31 & 132.64630° & 13.79319° \\
4 & p8 & 137.36386° & 20.28748° \\
4 & p16 & 136.59365° & 17.89575° \\
4 & p24 & 135.84396° & 15.50108° \\
4 & p32 & 135.11146° & 13.10394° \\
\hline
\end{tabular}
\end{table}

\begin{table}[htb!]
\centering
\caption{WFST L4 ETAs survey observing windows.}
\label{tab:WFST_obstime}
\begin{tabular}{cccc}
\hline
\hline
Day & Start Time & End Time\\
 & (UTC) & (UTC)\\
\hline
1 & 2025-09-22 21:21:36 & 2025-09-22 22:36:28\\
2 & 2025-09-23 21:18:37 & 2025-09-23 22:33:38\\
3 & 2025-09-25 21:25:42 & 2025-09-25 22:40:31\\
4 & 2025-09-29 21:28:04 & 2025-09-29 22:54:47\\
\hline
\end{tabular}
\end{table}

To schedule our survey for ETAs, we calculated the observability of the L4 point for WFST and found that the optimal observing window occurs between September and November each year, shortly before sunrise, due to favorable observing geometry.  
To avoid trailing of ETAs on the images, we set the exposure time to 30~s; for a typical proper motion of 1$^\circ$\,day$^{-1}$, an ETA moves approximately 1.25$''$ during a 30~s exposure—comparable to the seeing.  
All exposures were taken in the $r$ band, as asteroids are brighter in $r$ than in any other filter except $w$, based on limiting magnitudes \citep{leiLimitingMagnitudesWide2023} and color corrections \citep{wangHeliocentricorbitingObjectsProcessing2025}. The $w$ band was not used because template images in this filter were unavailable for image subtraction.

We defined a rectangular region centered on the L4 point as of 2025 September 25, spanning 20$^\circ$ in ecliptic longitude and 10$^\circ$ in ecliptic latitude, and allocated 32 ($4 \times 8$) pointings to cover the ``L4 cloud'' (see Fig.~\ref{fig:footprints_radec} and Table~\ref{tab:Targets_radec}). This region was chosen because stable ETAs can exhibit large libration amplitudes and low orbital inclinations \citep{yeagerMEGASIMDistributionDetection2024}. To date, this constitutes the largest dedicated ETAs survey, enabled by WFST's wide field of view.

We conducted four observing sessions as listed in Table~\ref{tab:WFST_obstime}. 
The right ascension and declination of the 32 pointings were kept fixed throughout the campaign. 
Although the L4 Lagrange point drifted by several degrees in ecliptic longitude over the observing window, our 20$^\circ \times 10^\circ$ survey area is sufficiently large to encompass the entire drift trajectory of the L4 region during this period.
Each session cycled through all 32 pointings three times, requiring approximately 80 minutes (including overheads) before nautical dawn. 
This cadence satisfies the requirements of both the MOPS-like ``single-night'' algorithm \citep{denneauPanSTARRSMovingObject2013} and the multi-night HelioLinC algorithm \citep{holmanHelioLinCNovelApproach2018}, thereby maximizing the recovery efficiency of moving objects. 
Due to altitude constraints near the horizon, we divided the 32 pointings into four groups. 
Observations were carried out by cycling through Group~1 three times, followed by Group~2, and so on for all groups.

The spatial footprint of our L4 ETAs survey is illustrated in Fig.~\ref{fig:footprints_radec}. The WFST focal plane comprises nine CCDs, which inherently introduce narrow inter-chip gaps. Excluding these gaps, the effective imaging area of each individual exposure is approximately octagonal, as the four corners of the CCD array lie outside the usable field of view and are thus truncated. Owing to a scheduling error during the observing run on 2025 September 22, only 20 of the 32 intended pointings were successfully executed. Of these, 12 fields were observed six times each instead of the nominal three visits. Observations conducted on 2025 September 22 and 25 were carried out under favorable weather conditions, yielding fully usable datasets. In contrast, poor weather on the nights of 2025 September 23 and 29 both degraded the limiting magnitudes and resulted in partial data loss. Specifically, for certain pointings on these nights, only a subset of the FITS files met the quality criteria of the WFST Data Reduction Pipeline \citep{cai25meterWideField2025}. Given the strict visit-time requirements outlined in Section~\ref{subsec:data_reduction}, this data incompleteness subsequently reduced the effective sky coverage available for asteroid searches.

\subsection{Data Reduction}\label{subsec:data_reduction}

The exposures were processed by the WFST Data Reduction Pipeline \citep{cai25meterWideField2025}.  
Each exposure was divided into 36 $4.6\mathrm{K} \times 4.6\mathrm{K}$ FITS image files, each representing a quadrant of a CCD.  
In addition to science images and source catalogs, difference images and associated catalogs in AVRO format are also produced.  
Both the science catalogs and the AVRO files are then used as input to the Heliocentric-orbiting Objects Processing System \citep[HOPS]{wangHeliocentricorbitingObjectsProcessing2025} for asteroid detection with WFST. We adopted a wide proper motion range of $0.02$--$1.5^\circ\,\mathrm{day}^{-1}$ to search for unknown solar system objects, not limited to ETAs. Following \citet{markwardtSearchL5Earth2020}, we assume that stable ETAs exhibit motions between $0.75$ and $1.25^\circ\,\mathrm{day}^{-1}$, a range that lies entirely within our adopted search interval.

For a blind search using HOPS, the single-night mode requires at least three observations on a given night, whereas the multi-night mode requires at least two observations per night over at least three nights within a 14-day window. Consequently, only regions with $\geq 3$ exposures per night—as shown in Fig.~\ref{fig:footprints_radec}—are suitable for an L4 ETA search in single-night mode. Although we also executed HOPS in multi-night mode, no new ETAs were identified. All detections corresponded to known asteroids, with one exception: an unknown candidate detected twice on September 23, which was subsequently identified as the known comet C/2022~QE78. This comet was also detected in single-night mode on September 25 and 29; it was not detected on September 22 because it did not fall within the observed regions. To derive a conservative upper limit on the L4 ETA population and for simplicity, we restrict our analysis to the results obtained from the single-night mode.

\section{Methods}

\subsection{Upper Limit Calculation}\label{subsec:upperlimit_cal}

The non-detection of ETAs in our survey can be modeled as a Poisson process. 
Following the single-pipeline counting experiment described in Section~2 of \citet{suttonUpperLimitsCounting2009}, the 95\% confidence upper limit on the number of ETAs is given by\footnote{In Section~2 of \citet{suttonUpperLimitsCounting2009}, the 90\% confidence upper limit for zero observed events is given as $2.30/\epsilon$. The 95\% version follows from the same Poisson logic, where the constant is $-\ln(0.05) \approx 2.997$, commonly rounded to 3. Equation~\ref{eq:U(H)=3/epsilon} is therefore a direct extension of Sutton's single-pipeline case to the 95\% confidence level.}:
\begin{equation}\label{eq:U(H)=3/epsilon}
    U(H) = \frac{3}{\epsilon(H)},
\end{equation}
where $\epsilon(H)$ is the total detection efficiency for an ETA of absolute magnitude $H$. 
This efficiency accounts for both the spatial coverage of our survey and the pipeline's recovery rate.

Let $C$ denote the coverage percentage of the stable L4 ETA by our observations (i.e., the 2-D spatial completeness), 
and let $R(H)$ be the recovery rate—the probability that our detection pipeline identifies an ETA of absolute magnitude $H$ within the observed fields. 
Assuming these factors are independent, the total detection efficiency is:
\begin{equation}\label{eq:epsilon=RxC}
    \epsilon(H) = R(H) \cdot C.
\end{equation}
Substituting into Equation~(1), the 95\% confidence upper limit becomes:
\begin{equation}
    U(H) = \frac{3}{R(H) \cdot C}.
\end{equation}

We estimate the recovery rate $R(V)$ using a catalog-based approach: 
the ratio of known asteroids recovered by our blind detection pipeline to those whose ephemerides fall within our survey footprint and satisfy our blind search constraints. 
Ephemerides are computed using the Python package \textsc{ALEPH} \citep{penazamudioALEPHFastAccurate2021} and the \texttt{MPCORB.DAT}\footnote{\url{www.minorplanetcenter.net/data}} file from the Minor Planet Center.

Although injection-recovery simulations are the standard method for quantifying detection efficiency, they are currently impractical for our pipeline due to computational and implementation constraints. 
Nevertheless, as detailed in Section~\ref{sec:results}, we demonstrate that our catalog-based estimate is both reasonable and applicable for ETA population studies. 

Since $R(V)$ depends on apparent magnitude while our population model requires $R(H)$ as a function of absolute magnitude $H$, we convert between $V$ and $H$ using the standard asteroid photometric relation:
\begin{equation}\label{eq:V-H}
    V = H + 5\log_{10}(r\Delta) - 2.5\log_{10}[\phi(\alpha)] + \delta,
\end{equation}
where $r$ and $\Delta$ are the heliocentric and geocentric distances (in au), $\alpha$ is the phase angle, $\phi(\alpha)$ is the phase function, and $\delta$ represents rotational lightcurve variations. 
For population-level efficiency estimates, we neglect $\delta$ under the assumption that rotation phases are randomly distributed, causing $\langle \delta \rangle \approx 0$.

We adopt the two-parameter $H$–$G$ magnitude system of \citet{bowellApplicationPhotometricModels1989} with $G = 0.15$ to compute $\phi(\alpha)$. 
For an idealized survey near the L4/L5 points covering a small sky area, we approximate $r \approx \Delta \approx 1$~au and $\alpha \approx 60^\circ$, yielding:
\begin{equation}\label{eq:V=H+2.15}
    V \approx H + 2.15.
\end{equation}
Under this approximation, the absolute-magnitude-dependent recovery rate is obtained via:
\begin{equation}
    R(H) = R(V = H + 2.15).
\end{equation}

\begin{figure}[ht!]
   \centering
   \includegraphics[width=\hsize]{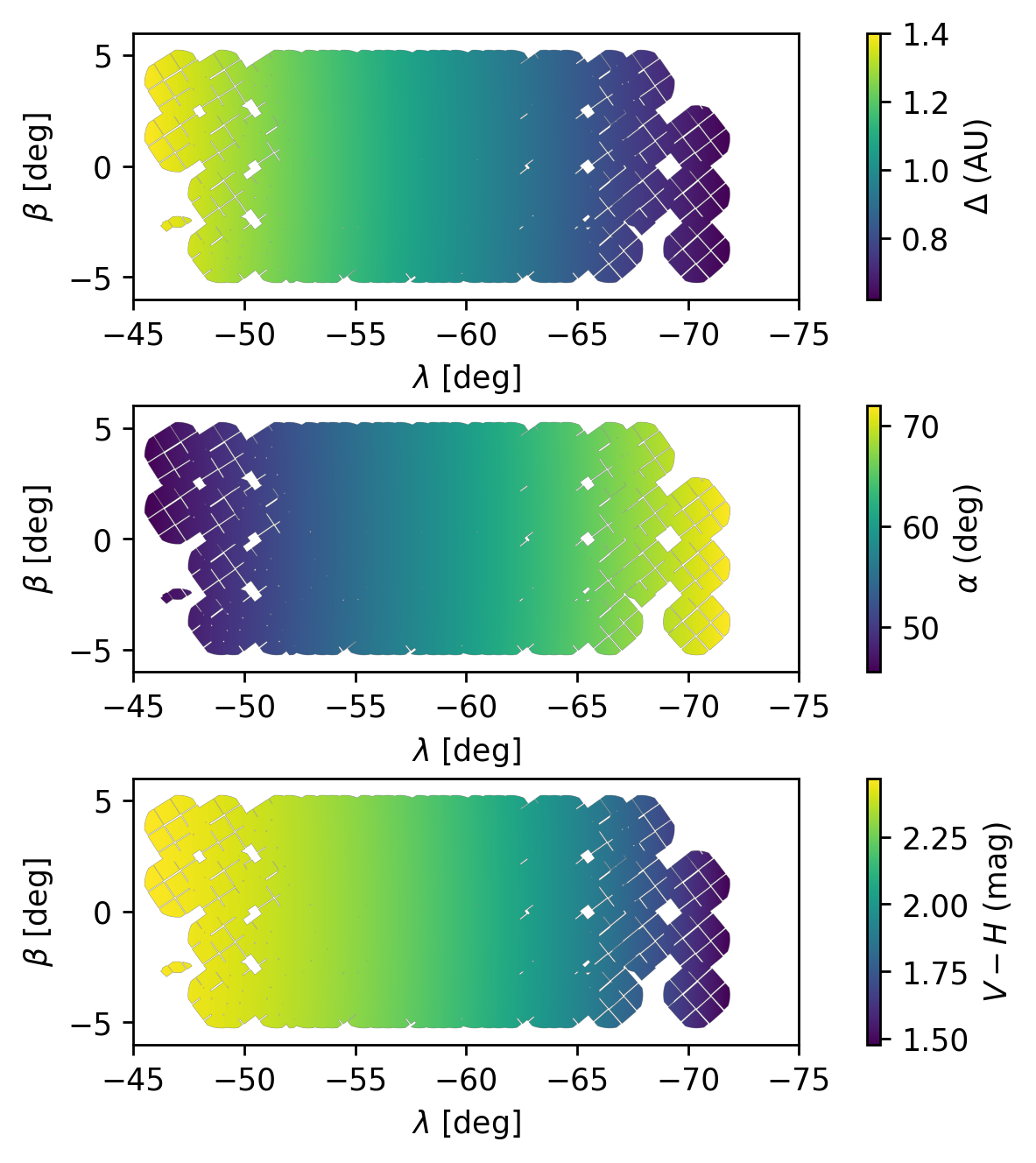}
   \caption{
       Geocentric distance $\Delta$, phase angle $\alpha$, and apparent-minus-absolute magnitude ($V - H$) for an ETA with heliocentric distance $r = 1$~au, computed across our L4 survey field. 
       The colored region indicates sky areas with at least three usable exposures on at least one of the four observing nights. 
       Coordinates $(\lambda, \beta)$ are geocentric ecliptic longitude and latitude in a Sun--Earth co-rotating frame, where the L4 point is stationary at $(-60^\circ, 0^\circ)$ and the Sun is fixed at $(0^\circ, 0^\circ)$.
   }
   \label{fig:V-H_variation}
\end{figure}

For surveys covering extended regions around L4 or L5, the simplified assumptions ($r = \Delta = 1$~au, $\alpha = 60^\circ$) introduce significant errors due to variations in viewing geometry. 
Instead, we fix only the heliocentric distance at $r = 1$~au and compute $\Delta$ and $\alpha$ individually for each sky position within the survey footprint.

Figure~\ref{fig:V-H_variation} illustrates the spatial variation of $\Delta$, $\alpha$, and $V - H$ across our L4 survey area. 
Smaller $\Delta$ and $\alpha$ enhance detectability (yielding brighter apparent magnitudes), but $\Delta$ dominates the variation in $V - H$ over our field. 
Critically, $V - H$ varies by up to $\sim$1~mag, highlighting the necessity of a spatially resolved detection efficiency model for wide-area ETA surveys.

Since detections in different sky regions are independent, the total detection efficiency $\epsilon(H)$ is obtained by integrating the local recovery probability weighted by the underlying ETA spatial density:
\begin{equation}\label{eq:epsilon_integration}
    \epsilon(H) = \iint R\big(V(H,\lambda,\beta)\big) \, D(\lambda,\beta) \cos\beta \; \mathrm{d}\lambda \, \mathrm{d}\beta,
\end{equation}
where $D(\lambda,\beta)$ is the normalized probability density of stable L4 ETAs (i.e., the dynamical model from numerical simulations), and the factor $\cos\beta$ accounts for spherical geometry on the celestial sphere. 
The total spatial completeness $C$ is then:
\begin{equation}\label{eq:C_and_D}
    C = \iint D(\lambda,\beta) \cos\beta \; \mathrm{d}\lambda \, \mathrm{d}\beta.
\end{equation}

For a multi-night survey comprising $K$ observing epochs, the total recovery probability at each sky position is given by the complement of failing to detect the object on all nights:
\begin{equation}
    R\big(V(H,\lambda,\beta)\big) = 1 - \prod_{k=1}^{K} \left[ 1 - R_k\big(V(H,\lambda,\beta)\big) \right],
\end{equation}
where $R_k$ denotes the single-night recovery rate on the $k$-th epoch, and detections across epochs are assumed statistically independent. In this work, we apply the formalism to a four-night survey, so $K = 4$.

In practice, we discretize the $(\lambda,\beta)$ plane into a grid and approximate the integral in Equation~\eqref{eq:epsilon_integration} as a sum:
\begin{equation}\label{eq:epsilon_sum}
    \epsilon(H) = \sum_{n=1}^{N} \left( 1 - \prod_{k=1}^{K} \left[ 1 - R_k\big(V_n(H)\big) \right] \right) D_n \, S_n,
\end{equation}
where $S_n$ is the solid angle subtended by the $n$-th grid cell, $D_n = D(\lambda_n, \beta_n)$ is the spatial probability density at that cell, and $V_n(H) = V(H, \lambda_n, \beta_n)$ is the apparent magnitude corresponding to absolute magnitude $H$ at position $(\lambda_n, \beta_n)$.

Finally, we clarify that our goal is not to derive a spatially resolved recovery rate $R_k(V)$. Rather, we treat $R_k(V)$ as a simple function that depends solely on $V$ for each night $k$. In other words, for a given night, we assume $R_k(V)$ is uniform across all surveyed regions. However, $R_k(H)$ is spatially resolved, as we aim to account for the spatial variation of $V-H$ in a wide-field survey. A spatially resolved $R_k(V)$ is unnecessary for both injection-recovery simulations and catalog-based estimates. In the former case, implementing such resolution would be prohibitively tedious, especially for large-field surveys, as it would require a substantial increase in the number of simulations. In the latter case, the lack of asteroids in certain regions would render recovery rate estimates infeasible. Since our objective is merely to obtain a reference recovery rate to estimate the upper limit on the ETA population, we adopt the catalog-based approach to compute an averaged $R_k(V)$ per night, which then provides an estimate of the ETA population.

\subsection{ETA Simulation}\label{subsec:eta_simulation}
We performed a series of simulations to investigate the spatial distribution of stable L4 ETAs. 
These simulations were carried out using the N-body integration package \textsc{REBOUND} \citep{reinREBOUNDOpensourceMultipurpose2012}, with the Sun and all eight planets included in the model.  
First, we initialized 10 million test particles and integrated their orbits for 10\,kyr using the IAS15 integrator \citep{reinIas15FastAdaptive2015}. 
The semimajor axes, eccentricities, and inclinations were randomly sampled from normal distributions—with means of 1~au, 0, and $0^\circ$, and standard deviations of 0.0667~au, 0.0667 (absolute values enforced), and $10^\circ$, respectively—while all angular orbital elements were drawn uniformly from $[0, 2\pi)$.  
This initial distribution matches that of \citet{lifsetSearchL4Earth2021}.  
During the integration, any particle that ventured beyond the half-space on the Earth-leading side was removed.  
At the end of this phase, 93\,795 particles survived.

We compared the orbital element distributions of the survivors with those of the initial ensemble and found no significant differences except in the semimajor axis. 
The surviving particles exhibit a tight clustering near 1.0~au, with a standard deviation of only 0.0013~au—much less than the initial spread of 0.0667~au; none deviate from 1.0~au by more than 0.0029~au. 
Even particles that survived only 100~yr show deviations smaller than 0.0065~au from 1.0~au.
We also examined the two known ETAs, (706765)~2010~TK$_7$ and (614689)~2020~XL$_5$, as listed in the Minor Planet Center (MPC) database; both lie within $1\sigma$ of 1.0~au.  
Together, these results indicate that—even for temporary ETAs—the semimajor axis is highly constrained, while other orbital elements are comparatively permissive.

\begin{figure}[ht!]
   \centering
   \includegraphics[width=\hsize]{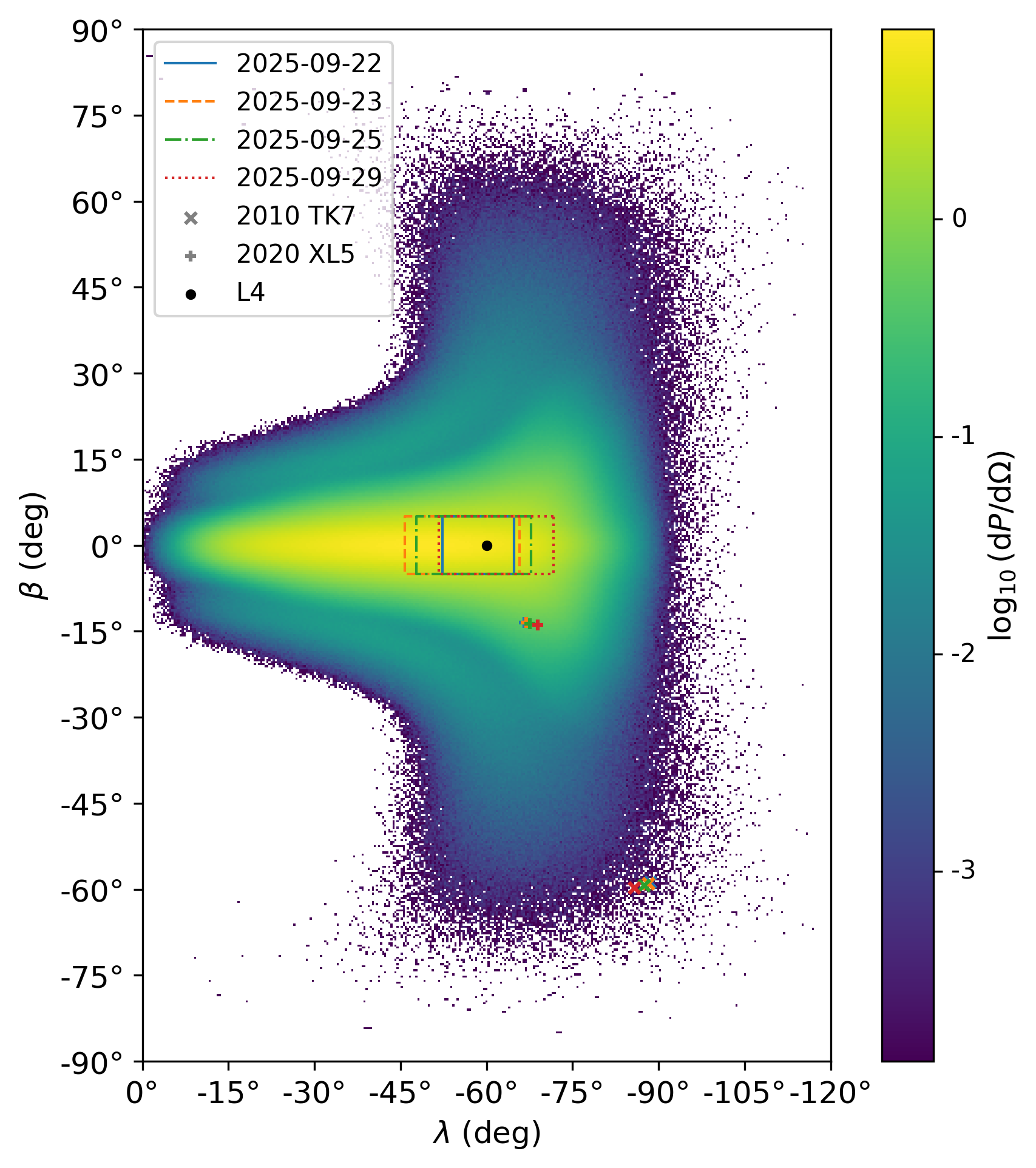}
   \caption{
       Surface probability density of stable L4 ETAs, expressed as 
       $D(\lambda,\beta) = \mathrm{d}P/\mathrm{d}\Omega$, 
       where $P$ is the total probability of finding an ETA and $\Omega$ denotes solid angle on the sky. 
       The color scale shows $\log_{10}\!\big(D(\lambda,\beta)\big)$, 
       with $D(\lambda,\beta)$ normalized such that $\iint D(\lambda,\beta) \cos\beta \, \mathrm{d}\lambda \, \mathrm{d}\beta = 1$. 
       This $D(\lambda,\beta)$ is the spatial probability density used in Equation~\eqref{eq:epsilon_integration}. 
       The black dot marks the L4 Lagrange point at $(-60^\circ, 0^\circ)$. 
       The four rectangles roughly indicate the sky coverage for each of the four observing nights. 
       The positions of the two known L4 ETAs are also marked: 
       (706765)~2010~TK$_7$ is shown with ``$\times$'' symbols and (614689)~2020~XL$_5$ with ``$+$'' symbols. 
       Their colors correspond to the same four nights as the rectangles, illustrating the motion of these objects over the survey period.
   }
   \label{fig:L4ET_density_map}
\end{figure}

We then divided the 93\,795 survivors into 100 subsets and propagated their orbits for 10~Myr in parallel using the \textsc{REBOUND} WHFast integrator \citep{reinWhfastFastUnbiased2015a}, with 17th-order symplectic correctors \citep{wisdomSymplecticMapsNbody1991}, following the setup of \citet{yeagerMEGASIMLifetimesResonances2022}.  
This switch from IAS15 was necessary because preliminary tests showed that IAS15 would be computationally prohibitive for such long integrations.  
We defined $2^{15} = 32\,768$ uniformly spaced checkpoints and, at each one, checked whether a particle had crossed the L$_3$ point or the Earth.  
After 10~Myr, 57\,261 particles remained and were classified as stable L4 ETAs.  
We computed their geocentric ecliptic coordinates in the Sun--Earth co-rotating frame as defined in Fig.~\ref{fig:V-H_variation}.  
The resulting spatial distribution (Fig.~\ref{fig:L4ET_density_map}) consists of two components: a high-density, horizontally oriented ``wizard hat'' and a more diffuse ``top hat'' background.  
We interpret these as corresponding to the two ETA populations with distinct inclination ranges mentioned in Section~\ref{Introduction}.  
Indeed, the inclination distribution of the 10~Myr ETAs exhibits a clear gap near $20^\circ$, separating the population into low- and high-inclination groups.  
Furthermore, when comparing the 93\,795 10~kyr survivors with the 57\,261 10~Myr ETAs, we find that the standard deviation of the semimajor axis remains essentially unchanged, whereas that of the eccentricity decreases by approximately 25\%. This dynamical filtering may be attributed to secular resonances.

\section{Results}\label{sec:results}

\subsection{Detections}\label{subsec:detections}
Our survey detected thousands of solar system objects, ranging from NEOs to comets. Table~\ref{tab:orbit_type} summarizes the daily detections by dynamical class. Since many objects were observed on multiple nights, we also report the number of unique objects (de-duplicated across epochs) in the final row. Several unknown asteroid candidates appear in our results, but these are either artifacts or comets (comet orbits are not included in our identification process). Thus, no new solar system objects were found.

We also examined the two known ETAs. During our survey, asteroid (706765)~2010~TK$_7$ was predicted to have an apparent magnitude of $V \sim 21.6$ mag, which is not too faint to be detectable. However, it is located near $(-87^\circ, -59^\circ)$ (September 25) in Fig.~\ref{fig:L4ET_density_map}, a very rare region where stable ETAs would reside. Most importantly, it lies in the southern sky, where the altitude is too low for our observations. Asteroid (614689)~2020~XL$_5$ is located near $(-68^\circ, -14^\circ)$ (September 25) in Fig.~\ref{fig:L4ET_density_map}, also outside our survey region. Moreover, its V-band magnitude was $\sim 23.3$ mag, too faint to be recovered by this survey even if it had been within the surveyed area.

\begin{table*}[htb!]
\centering
\caption{Orbit type of detected solar system objects}
\label{tab:orbit_type}
\begin{tabular}{lrrrrrrrrrrr}
\hline
\hline
Date & Aten & Apollo & Amor & MC & Hungaria & Phocaea & MBA & Hilda & DO & Comet & Total \\
\hline
2025-09-22 & 0 & 3 & 6 & 14 & 16 & 21 & 2661 & 13 & 2 & 2 & 2738 \\
2025-09-23 & 0 & 2 & 2 & 5 & 8 & 10 & 1370 & 7 & 0 & 2 & 1406 \\
2025-09-25 & 2 & 3 & 6 & 23 & 22 & 29 & 3648 & 11 & 1 & 3 & 3748 \\
2025-09-29 & 1 & 1 & 1 & 11 & 5 & 9 & 1553 & 5 & 0 & 1 & 1587 \\
\hline
\hline
Total & 3 & 9 & 15 & 53 & 51 & 69 & 9232 & 36 & 3 & 8 & 9479 \\
\hline
Unique & 3 & 7 & 8 & 35 & 33 & 38 & 5395 & 22 & 2 & 4 & 5547 \\
\hline
\end{tabular}

\footnotesize{
\textbf{Note:} MC = Mars-crosser, MBA = Main Belt Asteroid, DO = Distant object
}
\end{table*}

To assess potential biases in our recovery efficiency, we examined the dependence of the catalog-based recovery rate $R_k(V)$ on apparent sky motion. We find no significant correlation between recovery rate and proper motion across our search range, which encompasses the expected motion of stable ETAs ($0.75$--$1.25^\circ\,\mathrm{day}^{-1}$). This is further demonstrated by our successful recovery of fast-moving known asteroids. Table~\ref{tab:fast_sso} lists cataloged objects in our AVRO data with proper motion exceeding $0.65^\circ\,\mathrm{day}^{-1}$. Of the 15 entries, 12 were successfully recovered by our blind search pipeline. The three unrecovered cases correspond to objects that yielded only two detections in a single night—insufficient to satisfy the minimum three-detection requirement for tracklet formation. Notably, several recovered objects exhibit motions $\gtrsim 1^\circ\,\mathrm{day}^{-1}$, comparable to or exceeding typical stable ETA rates. Furthermore, if we restrict our analysis to objects with three or more detections on a single night in the AVRO data, all objects not recovered by our blind search exhibit proper motions below $0.6^\circ\,\mathrm{day}^{-1}$. This suggests that our blind search may be more sensitive to fast-moving objects.

These results confirm that our pipeline is capable of detecting and linking fast-moving asteroids, and that the recovery efficiency is not suppressed at high proper motions. Consequently, the catalog-based $R_k(V)$ defined in Section~\ref{subsec:upperlimit_cal} remains applicable for ETA population estimates.

\begin{table*}[htb!]
\centering
\caption{Solar system objects with proper motion > 0.65 deg/day in our AVRO data}
\label{tab:fast_sso}
\begin{tabular}{lcccccccc}
\hline
\hline
Designation & MJD & R.A. & Dec. & V & Type & N\_det & PM & Found  \\
 & & deg & deg & mag & & & deg/day & \\
\hline
(615555)& 60940.901609 & 122.506667 & 21.07854 & 22.35 & MC & 4 & 0.672 & True  \\
(89355)& 60943.933530 & 132.592881 & 16.84339 & 20.00 & Amor & 3 & 0.674 & True  \\
(468583)& 60940.907106 & 121.714833 & 16.02130 & 20.55 & Apollo & 6 & 0.684 & True  \\
(131702)& 60941.928206 & 133.197435 & 18.09701 & 19.14 & MC & 2 & 0.688 & False  \\
(131702)& 60943.932986 & 134.412363 & 17.31660 & 19.12 & MC & 2 & 0.689 & False  \\
(345722)& 60947.936713 & 134.591356 & 16.13681 & 20.17 & Aten & 6 & 0.691 & True  \\
(162741)& 60943.906157 & 125.269461 & 23.20511 & 20.90 & Amor & 3 & 0.697 & True  \\
(131702)& 60947.938924 & 136.820715 & 15.70592 & 19.08 & MC & 3 & 0.700 & True  \\
(162741)& 60940.905486 & 122.959868 & 23.25006 & 20.88 & Amor & 6 & 0.708 & True  \\
1994 WZ2& 60947.925775 & 132.146536 & 16.85806 & 21.32 & MC & 3 & 0.723 & True  \\
(217807)& 60941.888646 & 117.655879 & 23.10759 & 20.87 & Amor & 3 & 0.896 & True  \\
(507847)& 60940.918634 & 127.831080 & 22.67971 & 20.85 & Amor & 6 & 0.960 & True  \\
(507847)& 60943.919294 & 130.919367 & 22.15303 & 20.82 & Amor & 6 & 0.970 & True  \\
(140158)& 60941.889201 & 116.937270 & 18.87826 & 21.37 & Apollo & 2 & 1.093 & False  \\
(862524)& 60940.901609 & 120.945149 & 22.69763 & 21.40 & Apollo & 3 & 1.127 & True  \\
\hline
\end{tabular}

\footnotesize{
\textbf{Note:} N\_det = Number of detection, PM = Proper Motion, MC = Mars-crosser
}
\end{table*}

\subsection{L4 ETA Upper Limit}

\begin{figure}[ht!]
   \centering
   \includegraphics[width=\hsize]{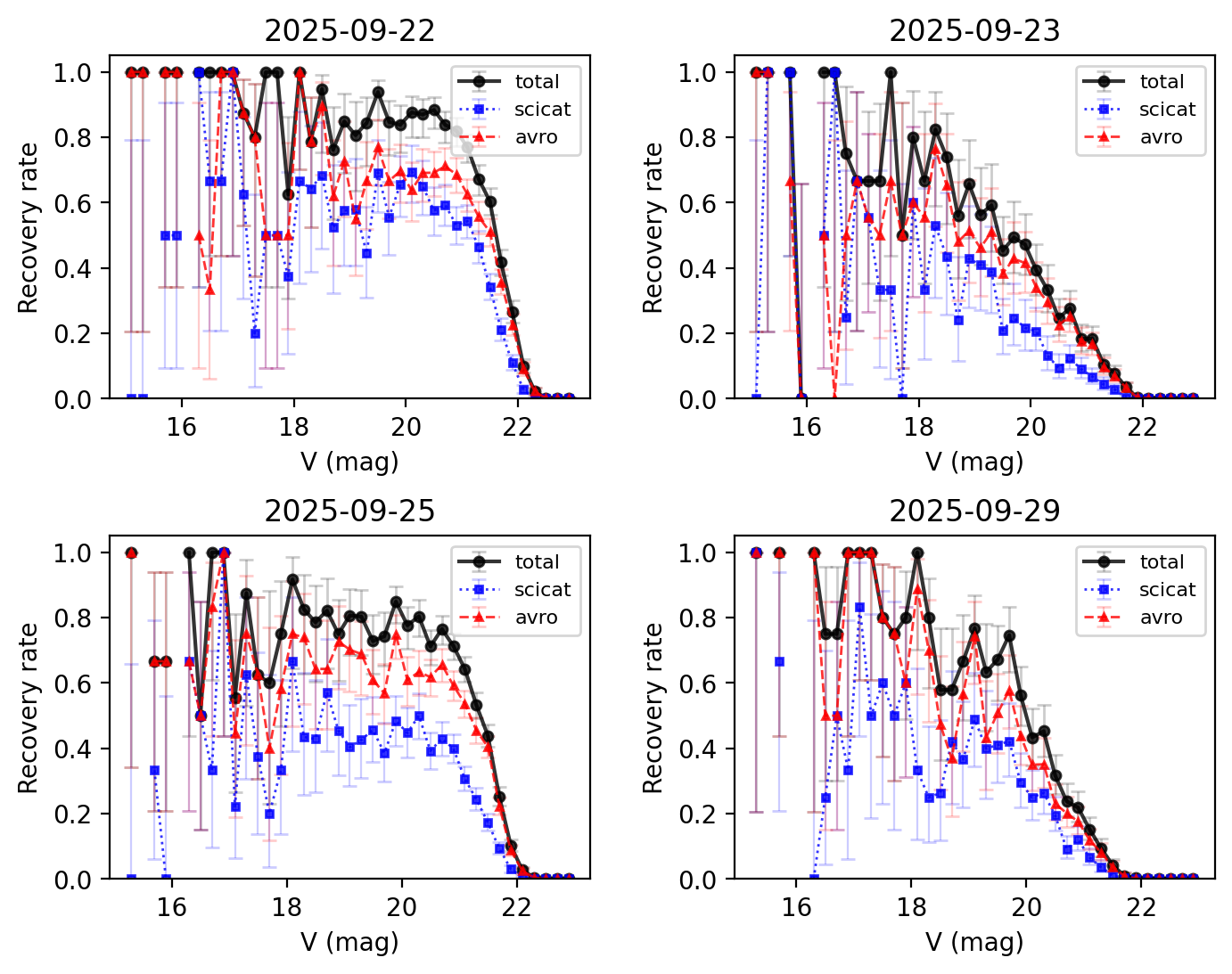}
    \caption{Recovery rate $R_k(V)$ of our L4 survey on each of the four nights. The red dashed line shows the recovery rate derived from AVRO data, the blue dotted line from the science catalog, and the black solid line the combined total recovery rate. Error bars are 95\% Wilson confidence intervals.}
    \label{fig:Recovery_vs_Vmag}
\end{figure}

The recovery rates for each night are shown in Fig.~\ref{fig:Recovery_vs_Vmag}. Error bars in Fig.~\ref{fig:Recovery_vs_Vmag} represent 95\% Wilson score confidence intervals for the binomial recovery proportion \citep{wilsonProbableInferenceLaw1927}. Due to poor observing conditions, the recovery rate $R_k(V)$ on 2025 September 23 and 29 falls below 50\% for $V \gtrsim 20$. In contrast, on September 22 and 25, the total recovery rate remains above 50\% up to $V \lesssim 21.5$. For $V < 18$, the recovery rate exhibits large fluctuations due to the small number of cataloged asteroids in this brightness range; the detection or non-detection of a single object can cause $R_k(V)$ to vary abruptly between 0 and 1.

The effective sky area available for ETAs analysis was 106.28, 134.66, 168.76, and 120.22~deg$^2$ on the four nights, respectively. For each night, we computed the fractional coverage as the integral of the long-term orbital stability probability density (see Fig.~\ref{fig:L4ET_density_map}) over the observed footprint, yielding values of 15.68\%, 19.66\%, 24.66\%, and 16.59\%, respectively. Accounting for overlapping regions between nights, the total unique surveyed area is 236.74~deg$^2$, corresponding to a cumulative probability coverage fraction of 33.24\%. This represents the largest dedicated survey for ETAs to date.

In this work, we defined a sky grid covering longitude from $-45^\circ$ to $-75^\circ$ and latitude from $-6^\circ$ to $6^\circ$, with a resolution of $l = 1/180^\circ$ in both dimensions, resulting in a $5400 \times 2160$ grid that encompasses our survey area (regions not usable or not covered are shown in white in Fig.~\ref{fig:V-H_variation}). For each grid cell, the solid angle is given by $S_n = l^2 \cos \beta = (\cos \beta) / 180^2 \deg^2$. Using bilinear interpolation, we derived the probability density $D_n$ (normalized to $\deg^{-2}$) for each grid cell, as displayed in Fig.~\ref{fig:L4ET_density_map}. For the missing values on the left side of Fig.~\ref{fig:Recovery_vs_Vmag}, we applied linear interpolation to obtain $R_k(V)$. In the subsequent calculations, we employed only the total recovery rate, rather than the AVRO or science catalog recovery rates.

\begin{figure}[ht!]
   \centering
   \includegraphics[width=\hsize]{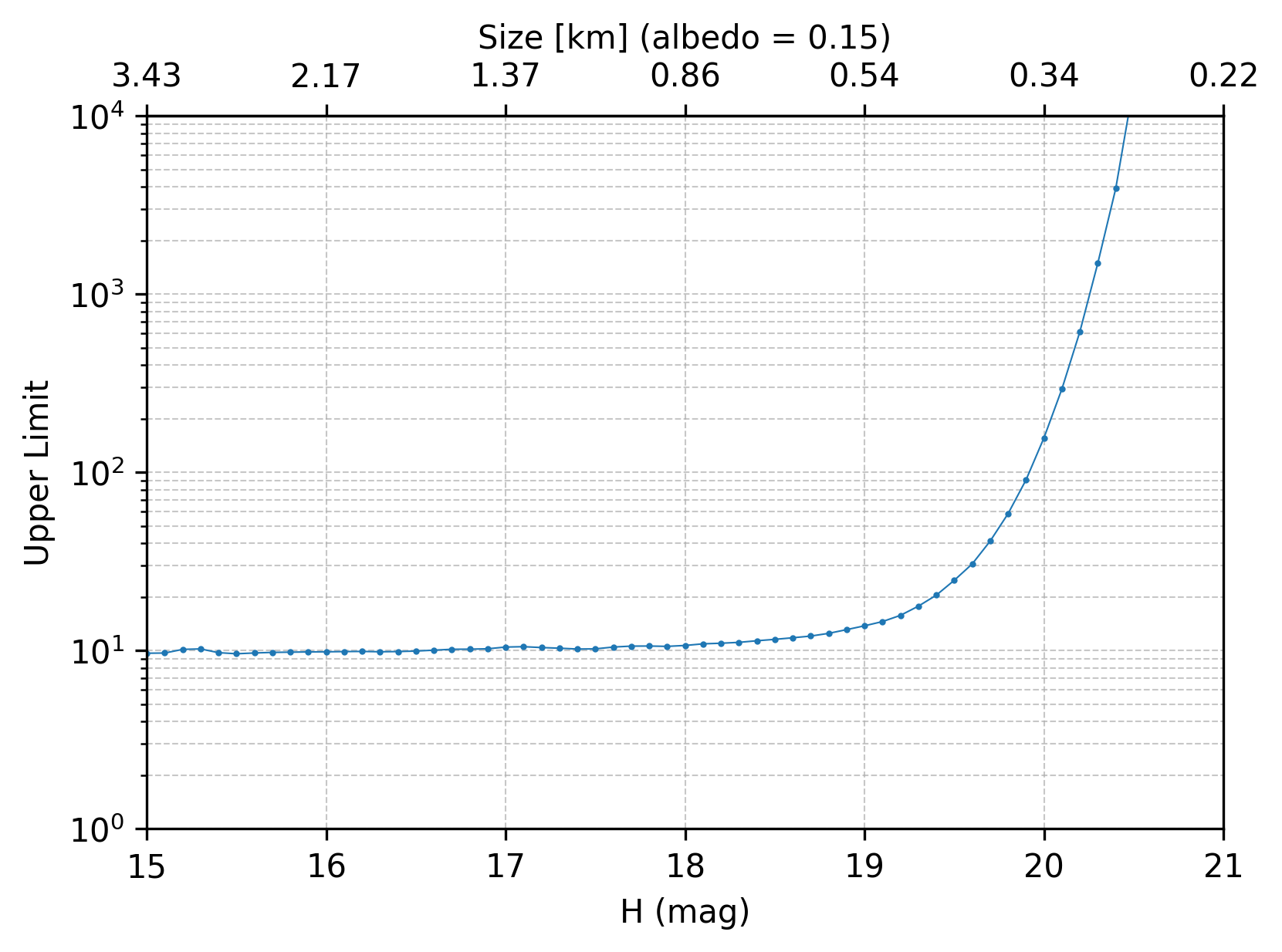}
    \caption{Differential upper limit $U(H)$ on the population of L4 ETAs, expressed as the maximum number per unit absolute magnitude, computed directly from our survey data using Equations~\eqref{eq:U(H)=3/epsilon} and~\eqref{eq:epsilon_sum}. The flat segment at bright magnitudes ($H \lesssim 19.5$) is limited by geometric coverage, while the steep decline at faint magnitudes ($H \gtrsim 20.5$) is driven by the survey’s limiting magnitude.
    }
    \label{fig:U_vs_H}
\end{figure}

Now, we can compute the upper limit $U(H)$ using Equations~\eqref{eq:epsilon_sum} and~\eqref{eq:U(H)=3/epsilon}, as shown in Fig.~\ref{fig:U_vs_H}. In this work, we assume an albedo of $0.15$ to convert $H$ to size, and the corresponding size scale is also shown in Fig.~\ref{fig:U_vs_H}. The integration in Equation~\eqref{eq:epsilon_sum} effectively smooths the $U(H)$ curve despite the noisy behavior of $R_k(V)$ at bright magnitudes (cf. Fig.~\ref{fig:Recovery_vs_Vmag}), enabling a high-resolution estimate of the population upper limit. For $H < 19.4$, our results yield $U(H) \lesssim 20$.

\section{Discussion}

We follow \citet{lifsetSearchL4Earth2021}'s approach to calculate the cumulative upper limit $N(<H)$ of the ETA population. We compute the spatial probability distribution using the simulation method described in Section~\ref{subsec:eta_simulation} to avoid overestimating the effective observational coverage. The cumulative upper limit $N(<H)$ is modeled as a broken power law:
\begin{equation}\label{eq:N_lt_H}
N(<H) = A \cdot 10^{\alpha H},
\end{equation}
where the slope $\alpha$ adopts values from the NEO size--frequency distribution model of \citet{schunova-lillySizefrequencyDistribution132017a}: $\alpha = 0.48 \pm 0.02$ for $13 < H < 16$, $\alpha = 0.33 \pm 0.01$ for $16 < H < 22$, and $\alpha = 0.62 \pm 0.03$ for $H > 22$.

Differentiating both sides of Equation~\eqref{eq:N_lt_H} with respect to $H$ yields the differential upper limit:
\begin{equation}\label{eq:dN_dH}
U(H) = \frac{dN}{dH} = A \cdot \alpha \cdot \ln(10) \cdot 10^{\alpha H}.
\end{equation}
Rearranging this expression gives:
\begin{equation}\label{eq:A_from_U}
A = \frac{U(H)}{\alpha \cdot \ln(10) \cdot 10^{\alpha H}}.
\end{equation}
To obtain the tightest possible cumulative upper limit, we minimize $A$ by evaluating Equation~\eqref{eq:A_from_U} across the range $16 < H < 22$. The minimum occurs at $H = 19.1$, where $U(H) \approx 14.5$ and $N(<H) \approx 19.1$.

\begin{figure}[ht!]
    \centering
    \includegraphics[width=\hsize]{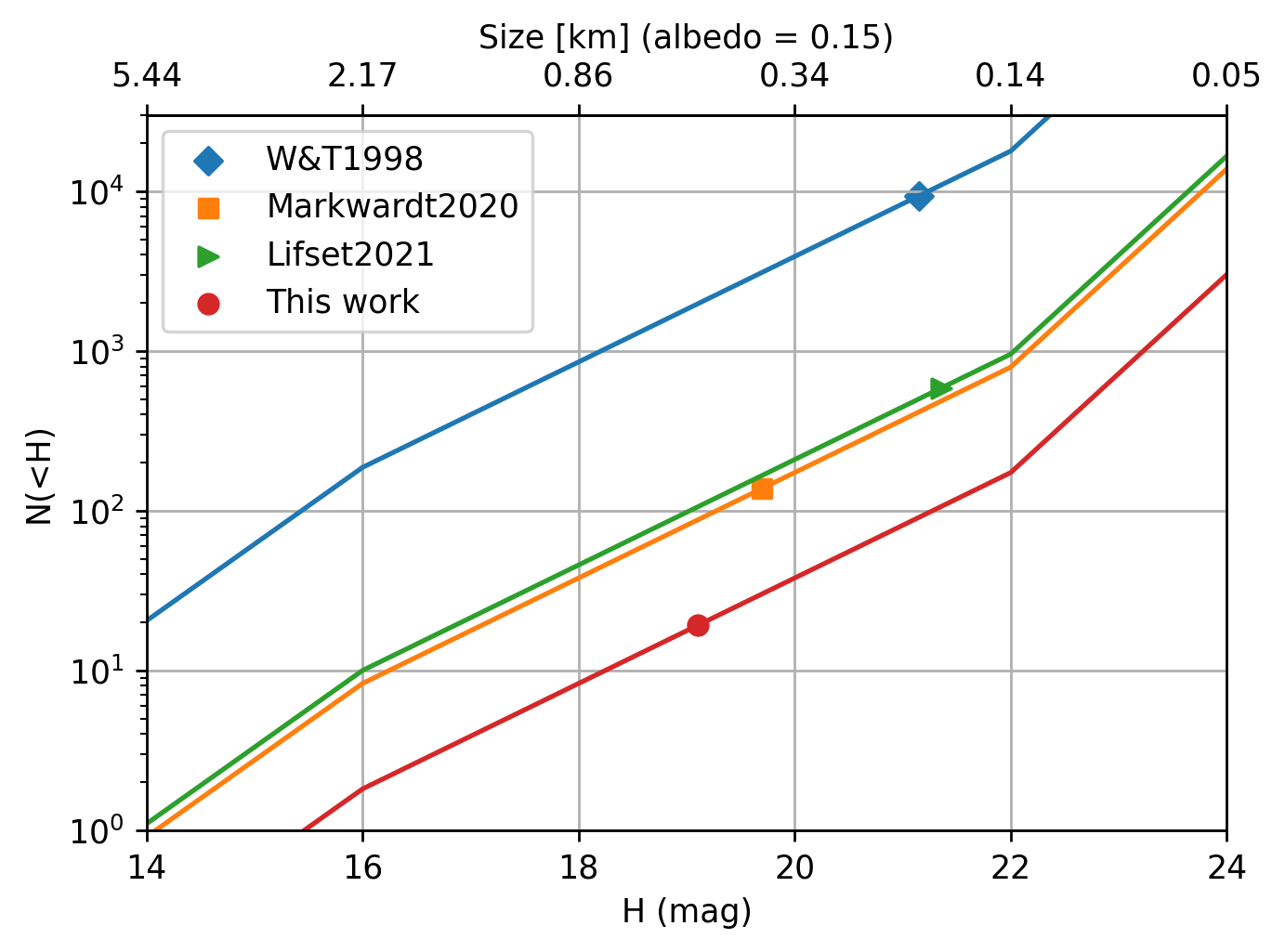}
    \caption{Cumulative upper limits on the L4/L5 ETA population from this work and previous surveys. The blue line and diamond show the result of \citet{whiteleyCCDSearchLagrangian1998}, gold line and square show the result of \citet{markwardtSearchL5Earth2020}, green line and triangle show the result of \citet{lifsetSearchL4Earth2021}, and red line and circle show the result of this work (with $N(H<19.1) \approx 19$). Results from \citet{cambioniUpperLimitEarths2018} and \citet{yoshikawaMissionStatusHayabusa22018a} are not plotted due to insufficient observational details, but their limits are weaker than that of \citet{markwardtSearchL5Earth2020}.
    }
    \label{fig:CumulativeUpperLimit}
\end{figure}

We compare our results with previous surveys in Fig.~\ref{fig:CumulativeUpperLimit}, assuming that the L4 and L5 regions are symmetric. For reference, our best-fit broken-power-law model yields $N(H<16) = 1.81$ and $N(H<22) = 173$. This represents the most stringent upper limit on the ETA population to date.

Current research on ETAs typically assumes that the L4 and L5 regions are symmetric. This assumption is not problematic, given that the paucity of detected ETAs precludes statistically conclusive results. Nevertheless, the two known ETAs are both located near L4. This raises the question of whether this observational bias is merely a coincidence or whether it hints at hidden clues regarding the evolutionary history of our Solar System. Notably, the number of L4 Jupiter Trojans exceeds that of L5 Jupiter Trojans, even after accounting for possible selection effects and faint completeness limits \citep{szaboPropertiesJovianTrojan2007}. This asymmetry suggests that Jupiter may have experienced a dynamical event—such as a scattering encounter with an ice giant \citep{nesvornyCAPTURETROJANSJUMPING2013}, rapid inward migration \citep{piraniConsequencesPlanetaryMigration2019}, or rapid outward migration \citep{liAsymmetryNumberL42023a}. Therefore, obtaining more comprehensive information on both L4 and L5 ETAs is valuable for understanding the dynamical history of the Earth and the broader Solar System.

As in previous ETA surveys, no stable ETAs were discovered in this work. One possible explanation is that, for ground-based telescopes, the observational windows for the L4 and L5 regions are limited to specific months or weeks, and observations must be carried out shortly after sunset or before sunrise at low altitudes—for instance, during twilight—which is far from ideal for fully utilizing a telescope's capabilities. For regions very close to the Sun, the altitude and observing window become too poor to yield useful data. Another possibility is that stable ETAs are both rare and small, or that none have survived to the present day. Dynamical studies incorporating the Yarkovsky effect suggest that small primordial ETAs are unlikely to have survived to the present day \citep{zhouOrbitalStabilityEarth2019a}. Our wide-area survey places a cumulative upper limit of $N(H<21.3) \lesssim 100$ on the population of large stable ETAs. Thus, the primordial ETA population is likely extremely scarce, if it exists at all. However, many temporary ETAs may still await discovery, as they could be spread over a wider region around the Lagrange points and exhibit a broader proper motion distribution.

If any primordial ETAs do exist, they likely reside in sky regions close to the Sun, where ground-based observations are severely constrained by low solar elongation, bright twilight conditions, and short observing windows near the horizon. Furthermore, the physical properties of ETAs remain poorly constrained; should their size–frequency distribution differ significantly from that of NEOs, current population estimates could be substantial underestimates.

To further constrain the ETA population, dedicated future surveys will be essential—particularly the Vera C.~Rubin Observatory's Legacy Survey of Space and Time \citep{jonesLargeSynopticSurvey2018} and space-based missions such as NASA's NEO Surveyor \citep{mainzerNearEarthObjectSurveyor2023}. The former achieves a deep limiting magnitude and simultaneously covers a large sky area, both of which are advantageous for discovering new ETAs. The latter, situated at the Earth–Sun L1 point, suffers no observing window constraints, is unaffected by adverse weather, and provides infrared observing capabilities that are more sensitive to asteroids—offering numerous advantages over ground-based telescopes for ETA discovery.

\newpage 

\begin{acknowledgements}
This work is supported by National Key Research and Development Program of China (2023YFA1608100). The authors gratefully acknowledge the support provided by the National Natural Science Foundation of China (NSFC, Grant No. 12233008), the CAS Project for Young Scientists in Basic Research (No. YSBR-092), Guizhou Provincial Major Scientific and Technological Programs XKBF (2025)010 and XKBF (2025)011, Cyrus Chun Ying Tang Foundations and the 111 Project for ``Observational and Theoretical Research on Dark Matter and Dark Energy'' (B23042). 
\end{acknowledgements}

\bibliographystyle{aa}
\bibliography{L4ETA}
\end{document}